\documentclass[%
 reprint,
superscriptaddress,
nofootinbib,
 amsmath,amssymb,
pra,
longbibliography
]{revtex4-1}
\usepackage[utf8]{inputenc}
\usepackage{amsmath,amsthm}
\usepackage{dcolumn}
\usepackage{bm}
\usepackage[colorlinks=true,citecolor=blue,urlcolor=blue]{hyperref}
\usepackage[pdftex]{graphicx}
\usepackage{braket}
\usepackage{color}
\usepackage[usenames,dvipsnames,svgnames,table]{xcolor}

\usepackage{amsmath}
\usepackage{latexsym}
\usepackage{tabularx, booktabs}
\usepackage{amssymb}
\usepackage{graphics,epstopdf}
\usepackage{graphicx}

\usepackage{soul}
\usepackage{graphicx}
\usepackage{amsfonts}
\usepackage{color}
\usepackage{caption}
\usepackage{subcaption}
\usepackage{wrapfig, framed, caption}
\usepackage{lipsum}

\begin{document}

\title{\textbf{Environmental effects on nonlocal correlations} }

\author{Tamal Guha}
\email{g.tamal91@gmail.com }
\affiliation{Physics and Applied Mathematics Unit, Indian Statistical Institute, 203 B. T. Road, Kolkata- 700108, India.}

\author{Bihalan Bhattacharya}
\affiliation{S. N. Bose National Centre for Basic Sciences, Block JD, Sector III, Salt Lake, Kolkata-700098, India.}

\author{Debarshi Das}
\email{debarshidas@jcbose.ac.in}
\affiliation{Center for Astroparticle Physics and Space Science (CAPSS), Bose Institute, Block EN, Sector V, Salt Lake, Kolkata-700091, India}

\author{Some Sankar Bhattacharya}
\email{somesankar@gmail.com}
\affiliation{Department of Computer Science, The University of Hong Kong, Pokfulam Road, Hong Kong.}

\author{Amit Mukherjee}
\email{amitisiphys@gmail.com}
\affiliation{Optics and Quantum Information Group, The Institute of Mathematical Sciences, HBNI, C.I.T. Campus, Taramani, Chennai-600113, India.}

\author{Arup Roy}
\email{arup145.roy@gmail.com}
\affiliation{S. N. Bose National Centre for Basic Sciences, Block JD, Sector III, Salt Lake, Kolkata-700098, India.}

\author{Kaushiki Mukherjee}
\email{kaushiki_mukherjee@rediffmail.com}
\affiliation{Department of Mathematics, Government Girls General Degree College, Ekbalpore, Kolkata-700023, India.}

\author{Nirman Ganguly}
\email{nirmanganguly@gmail.com}
\affiliation{ Department of Mathematics, Birla Institute of Technology and Science Pilani, Hyderabad Campus, Telengana-500078, India.}

\author{A. S. Majumdar}
\email{archan@bose.res.in}
\affiliation{S. N. Bose National Centre for Basic Sciences, Block JD, Sector III, Salt Lake, Kolkata-700098, India.}
\begin{abstract}

Environmental interactions are ubiquitous in practical instances of any quantum information processing protocol. The interaction results in depletion of various quantum resources and even complete loss in numerous situations. Nonlocality, which is one particular quantum resource marking a significant departure of quantum mechanics from classical mechanics, meets the same fate. In the present work we study the decay in nonlocality to the extent of the output state admitting a local hidden state model. Using some fundamental quantum channels we also demonstrate the complete decay in the resources in the purview of the Bell-CHSH inequality and a 3-settings steering inequality. We also obtain bounds on the parameter of the depolarizing map for which it becomes steerability breaking pertaining to a general class of two qubit states. 
\end{abstract}

\maketitle

\section{Introduction}

Nonlocality \cite{nll} is one of the key  features that sets quantum mechanics (QM) apart from classical mechanics. Quantum nonlocality is generally interpreted as the failure to describe  quantum mechanical (QM) correlations arising due to spacelike separated local quantum measurements on subsystems of a composite system by local realist models. Bell inequalities \cite{Bell} are used to reveal such incompatibility between QM and local-realism. A state which satisfies a set of Bell inequalities cannot be guaranteed as local, as there may exist another set of Bell  inequalities that it violates. On the other hand, QM violation of any Bell-CHSH inequality is a signature of quantum nonlocality \cite{chsh}. However, the complete set of Bell-CHSH inequalities is the necessary and sufficient criteria for local-realism in the $2 - 2 - 2$ experimental scenario ($2$ parties, $2$ measurement settings per party, $2$ outcomes 
 per measurement setting). A state is termed as local only if the correlations arising by performing local quantum measurements on it admit a local hidden variable (LHV) model \cite{lhv}.

The pioneering study by Einstein, Podolsky and Rosen (EPR) \cite{epr} demonstrating the incompleteness of the QM description of `reality' motivated Schrodinger to propose the concept of `quantum steering' \cite{scro}. EPR steering arises in the scenario where local quantum measurements on one part of a bipartite spatially separated system allow to prepare different ensembles on the other part. This scenario demonstrates EPR steering if these ensembles cannot be explained by a local hidden state (LHS) model \cite{steer1, steer2}. This kind of interpretation of steering has induced great interest in foundational research in recent times as evidenced by a wide range of studies \cite{st3, st4, st5, st6, st7, st8, nongauss, fgur1, fgur2, steer9, steer10}.  Reid first proposed a criterion for testing EPR-steering in  continuous-variable systems based on position-momentum uncertainty relation \cite{st9}, which was experimentally tested by Ou et al. \cite{st10}. More recently, Cavalcanti et al. constructed experimental EPR-steering criteria based on the assumption of the existence of LHS model \cite{st11}. This general construction is applicable to discrete as well as continuous-variable observables and Reid's criterion appears as a special case of this general formulation.

More importantly, entanglement \cite{ent} is a necessary condition for the demonstration Bell-nonlocality or steerability. The inequivalence of entanglement and nonlocality is exemplified by showing the existence of certain entangled states producing QM correlations that admit a LHV model \cite{ent2}. Bell-nonlocal states form a strict subset of steerable states which also form a strict subset of entangled states \cite{s1, steer1}.

Apart from being important candidates in foundational studies of QM, quantum nonlocal, steerable and entangled states serve as resource in various quantum information processing tasks, for example, teleportation \cite{tp}, randomness certification \cite{rc, rc2, rc3}, cryptography \cite{cg} and so on. Motivated by this fact a number of studies have been performed towards revealing `hidden' nonlocality from quantum states that failed to demonstrate nonlocality under the standard Bell scenario \cite{nll}.  Local filtering operation is one such procedure \cite{hnl1, hnl2}, which can be broadly classified into two categories (a) performing single local measurement \cite{hnl2} and (b) subjecting the state to suitable sequence of local measurements \cite{hnl1}. Similarly, the issue of revealing `hidden' quantum steerability by using local filters has also been studied \cite{hs1, s1}.

In practical scenarios, a state is subjected to ubiquitous environmental interaction and hence may lose its entanglement or nonlocal character partially or completely. Thus, it is of considerable interest to study the behaviour of entangled states as well as nonlocal states under local noise. The issue related to entanglement breaking channels which transform an entangled state into a separable one, has acquired a lot of significance as witnessed by a number of studies \cite{ebc2, ebc1}. On the other hand, studies on nonlocality breaking maps, which transform a Bell-CHSH nonlocal state into a local one, have also probed the role of environment in destroying quantum resources \cite{nbc}. Recently, a
general framework for analyzing resource theories based on resource destroying maps has been proposed \cite{resloyd}. In this context, one may also consider the case of incompatibility breaking maps \cite{incomp}, as incompatibility of quantum measurements is an important resource in quantum information. The study of such maps is intriguing as steerability and incompatibility have a one to one correspondence \cite{guhnesteering}. 

A separable state can be transformed into an entangled one when subjected to a global unitary action acting on the composite system. However, it has been shown that there are separable states, dubbed as absolutely separable states, which cannot be transformed into an entangled one under any global unitary interaction \cite{ass1, ass2, ass3, ass4}. Recently, the effect of global unitary interactions on the nonlocality of a state has been probed, with the focus on the Bell-CHSH inequality for two qubit systems. A state initially satisfying the Bell-CHSH inequality can violate it after a global unitary interaction. On the other hand, it has been demonstrated that there are states which preserve their Bell-CHSH local character under arbitrary global unitary action. These states are termed as absolutely Bell-CHSH local states \cite{abl1, abl2}. The question of transforming a separable, but not absolutely separable, state into an absolutely separable one under environmental interactions is important in practical situations and has been investigated recently~\cite{assd}. In the context of absolutely Bell-CHSH local states, the issue of transforming an absolutely Bell-CHSH local state into a nonlocal one has also been presented in a recent study \cite{abl3}. In a similar spirit, the effect of global unitary interactions on states demonstrating EPR steering has also been studied \cite{as1} in the context of steering inequalities derived by Cavalcanti et al. \cite{st11}. In particular, the issue of non-violation of the steering inequality with three measurement settings per party \cite{st11} by any two qubit system under arbitrary global unitary action has been studied in details. For our convenience, we will denote the states which preserve their non-violation of the steering inequality with three measurement settings per party derived by Cavalcanti et al.~\cite{st11} under arbitrary global unitary action as ``absolutely 3-settings unsteerable states". Studies have shown that any pure state cannot be absolutely Bell-CHSH local, or absolutely 3-setting unsteerable, or absolutely separable. However, one must remember that environmental interactions bound us to work with mixed states where such phenomena are practically very possible. Therefore, one can also understand the significance of such confrontations even through the lens of quantum information processing protocols. 

A natural question arising in this context is what are the possible instances that one should avoid while working with nonlocal resources under the influence of environment. In the present work we provide such illustrations with commonly used quantum channels. Our study is done in the purview of Bell-CHSH inequality and the 3-settings steering inequality mentioned in the text. The resultant states do not violate the inequalities even after consuming the power of global unitary operations. This is where the present study departs from previous works as it probes the feature that nonlocality may be non-retrievable even with global unitary operations.

On the foundational significance of our work, we have been able to generate entangled states which admit local hidden state models from initially nonlocal state. Werner, in his seminal paper \cite{ent2} had proved the existence of entangled states having LHV. That is,  entangled states exist which cannot give rise to nonlocality under arbitrary (non-sequential) measurements. There is a recent spurt in research in constructing entangled states having LHV or LHS model \footnote{Alice and Bob share a quantum state. Alice performs a measurement $x$ $\in$ $\{x_0, x_1, x_2, ..., x_n\}$ and obtains an outcome $a$ $\in$ $\{0, 1, 2, ..., d_A\}$ and Bob performs a measurement $y$ $\in$ $\{y_0, y_1, y_2, ..., y_m\}$ and obtains an outcome $b$ $\in$ $\{0, 1, 2, ..., d_B\}$. If they repeat these trials many times, they generate a correlation $p(ab\vert xy)$. The joint correlation is said to have LHV model if and only if $p(ab\vert xy)= \sum_\lambda \rho(\lambda) p(a \vert x,\lambda) p(b \vert y, \lambda)$, where $p(a \vert x,\lambda)$ and $p(b \vert y, \lambda)$ are arbitrary distributions conditioned on LHV $\lambda$. On the other hand, the joint correlation is said to have LHS model if and only if $p(ab\vert xy)= \sum_\lambda \rho(\lambda) p(a \vert x,\lambda) p(b \vert y, \sigma_\lambda)$, where $p(a \vert x,\lambda)$ is arbitrary distribution conditioned on LHV $\lambda$, $p(b \vert y, \sigma_\lambda)$ is the quantum probability of obtaining the outcome $b$ when the measurement $y$ is performed on the local hidden state $\sigma_\lambda$ (which is a quantum state).} \cite{caval,brunner} using semi-definite programming. In our present submission we have been able to generate an entangled state admitting LHS model due to environmental influence.  

We would like to note here that pure entangled states violate a suitably chosen Bell's inequality. However, purity is fragile under environmental influences and thus, mixed states are  more common in experimental scenarios. Mixed entangled states can have a local hidden variable model, the most celebrated example being of the Werner state in two qubits. Therefore, our present work deals with mixed entangled states.

We have studied the context of steerability breaking maps. Particularly we have obtained bounds on the parameter of the depolarizing map such that it becomes steerability breaking pertaining to a general class of two qubit states. This is particularly significant because steerability is a weaker form of nonlocality as compared to Bell nonlocality. 

The paper has been arranged in the following way. Starting from some preliminary discussions in section II we have illustrated the scenarios in section III. In section IV we have studied the transformation of Bell-CHSH nonlocal state to Bell-CHSH local, absolutely Bell-CHSH local and absolutely 3-settings unsteerable states under environmental interactions, followed by one of the  highlights of our analysis, {\it viz.}, generation of entangled states admitting LHS model from initially nonlocal state using two different quantum channels in section V. In section VI, we discuss on the analysis of steerability breaking channels in the context of the depolarizing map which is another interesting attribute  of our approach. Finally,  we conclude along with a discussion in section VII.
\section{Preliminaries}

Let us start with some preliminary ideas required for the present study.

\subsection{ Bell-CHSH locality }

A bipartite state is said to be Bell-CHSH local if and only if (iff) the correlations obtained by performing local quantum measurements on the two subsystems of the composite state (where the local measurements performed on one subsystem are spacelike separated from that on another subsystem) do not violate the Bell-CHSH inequality. The necessary and sufficient criteria for QM violation of the CHSH inequality by arbitrary bipartite qubit states has been established in \cite{hd}.\\

An arbitrary two qubit state can be expressed in terms of the Hilbert-Schmidt basis as
\begin{eqnarray}\label{rho}
\rho=\frac{1}{4}(\mathbb{I}\otimes\mathbb{I}+\vec{r}.\vec{\sigma}\otimes\mathbb{I}+\mathbb{I}\otimes\vec{s}.\vec{\sigma}+\sum_{i,j=1}^{3}t_{ij}\sigma_i\otimes\sigma_j).
\end{eqnarray}
Here $\mathbb{I}$ is the identity operator acting on $\mathbb{C}^2$; $\sigma_i$s are the three Pauli matrices; $\vec{r}, \vec{s}$ are vectors in $\mathbb{R}^3$ with norm less than or equal to unity; $\vec{r}.\vec{\sigma} = \sum_{i=1}^{3} r_i \sigma_i$ and $\vec{s}.\vec{\sigma} = \sum_{i=1}^{3} s_i \sigma_i$. The condition Tr$(\rho^2) \leq 1$ implies
\begin{equation}
\sum_{i=1}^{3} \Big( r_i^{2} + s_i^{2} \Big) + \sum_{i,j=1}^{3} t_{ij}^2 \leq 3,
\end{equation}
where the equality is achieved for the pure states. In addition, for being a valid density matrix, $\rho$ has to be positive semidefinite.

Let us consider the matrix $V=T^{\dagger}T$, where $T$ is the correlation matrix of the state (\ref{rho}) with matrix elements $t_{ij} = Tr( \rho \sigma_i \otimes \sigma_j )$. $u_1$, $u_2$ are two greatest eigenvalues of $V$. Let us consider the quantity given by,
\begin{eqnarray}
M(\rho)=u_1 + u_2.
\end{eqnarray}
The state given by Eq.(\ref{rho}) violates the Bell-CHSH inequality iff $M(\rho) > 1$. Hence, the state (\ref{rho}) is Bell-CHSH local iff $M(\rho) \leq 1$.

\subsection{Absolute Bell-CHSH locality }

The concept of absolutely Bell-CHSH local states has recently been introduced in \cite{abl1}. A Bell-CHSH local quantum state is said to be absolutely Bell-CHSH local  if the state remains Bell-CHSH local under the action of any global unitary operation.
If $a_1,a_2,a_3$ are the three largest eigenvalues of the given two qubit state $\rho$ taking in descending order, then the state $\rho$ is absolutely Bell-CHSH local \textit{iff} \cite{abl2}

\begin{equation}
\label{ablc}
A(\rho) = (2a_1+2a_2-1)^2+(2a_1+2a_3-1)^2 \leq 1.\\
\end{equation}

\subsection{Absolute 3-settings unsteerability}

Cavalcanti et al. have provided a series of steering inequalities to certify whether a bipartite state is steerable when each of the two parties are allowed to perform $n$ measurements on his or her part \cite{st11}. In particular for $n=3$, the inequality is given by,
\begin{equation}
\label{st3}
F^3 = \dfrac{1}{\sqrt{3}} \Big| \sum_{i=1}^{3} \langle A_i \otimes B_i \rangle \Big| \leq 1,
\end{equation}
where, $A_i= \hat{u_i}.\vec{\sigma}$, $B_i= \hat{v_i}.\vec{\sigma}$, $\vec{\sigma}=(\sigma_1, \sigma_2, \sigma_3)$ is a vector composed of Pauli matrices, $\hat{u_i} \in \mathbb{R}^3$ are unit vectors, $\hat{v_i} \in \mathbb{R}^3$ are orthonormal vectors. $\langle A_i \otimes B_i \rangle$ = Tr$(\rho  A_i \otimes B_i)$ with $\rho \in \mathcal{B}(\mathcal{H}_A \otimes \mathcal{H}_B)$ is the bipartite quantum system shared between the two parties.\\

The states, which preserve their non-violation of the steering inequality (\ref{st3}) under arbitrary global unitary action, are called ``absolutely 3-settings unsteerable states". A given two qubit state $\rho$ is absolutely 3-settings unsteerable \textit{iff} \cite{as1}
\begin{equation}
\label{ausc}
B(\rho) =3 \text{ Tr}(\rho^{2})-2(x_1 x_2+x_1 x_3 +x_1 x_4 +x_2 x_3 + x_2 x_4 +x_3 x_4) \leq 1,
\end{equation}
where $x_i$ are the eigenvalues of the two qubit state $\rho$.

\subsection{Quantum channels}

In practical scenarios it is very hard to isolate a qubit from its environment. Environmental interactions can be represented by different quantum channels. Quantum channels are completely positive trace preserving (CPTP) maps acting on the space of density matrices \cite{nc}. Every quantum channel admits the operator sum representation. Let $\varepsilon$ be a quantum channel, then its action on the state $\tilde{\rho}$ can be expressed as
\begin{equation}
\varepsilon(\tilde{\rho})=\sum_{i} K_i \tilde{\rho} K_{i}^{\dagger},
\end{equation}
where $K_i$'s are Kraus operators for the corresponding channel with $\sum_{i} K_{i}^{\dagger} K_i= \mathbb{I}$ ($\mathbb{I}$ is the identity operator). In the present study we restrict ourselves to four quantum channels, viz. 1) Phase-flip channel , 2)  Bit-flip channel, 3) Depolarizing channel, and 4)  Phase damping channel.

\subsubsection{ Phase-flip channel}

The action of the phase-flip channel on the state $\tilde{\rho}$ can be expressed as \cite{nc}
\begin{equation}
\varepsilon(\tilde{\rho})=\sum_i K_i \tilde{\rho} K_{i}^{\dagger}.
\end{equation}
The Kraus operators for the phase-flip channel are given by
\begin{center}

	$K_0=\sqrt{1-p} \mathbb{I}, K_1=\sqrt{p}\sigma_z$,
	
\end{center}
where $p$ is the channel strength with $0 \leq p \leq 1$. \\\\

\subsubsection{ Bit-flip channel}

The action of the bit-flip channel on the state $\tilde{\rho}$ can be expressed as \cite{nc}
\begin{equation}
\varepsilon(\tilde{\rho})=\sum_i K_i \tilde{\rho} K_{i}^{\dagger}.
\end{equation}
The Kraus operators for the bit-flip channel are given by
\begin{center}

	$K_0=\sqrt{1-p} \mathbb{I}, K_1=\sqrt{p}\sigma_x$
\end{center}
where  $p$ is the channel strength with $0 \leq p \leq 1$. 

\subsubsection{Depolarizing channel}

The action of the depolarizing channel on the state $\tilde{\rho}$ can be expressed as \cite{nc}
\begin{equation}
\varepsilon(\tilde{\rho})=\sum_i K_i \tilde{\rho} K_{i}^{\dagger}.
\end{equation}
The Kraus operators for the depolarizing channel are given by
\begin{center}
	$K_0=\sqrt{1-p} \mathbb{I}, K_1=\sqrt{\dfrac{p}{3}}\sigma_x,K_2=\sqrt{\dfrac{p}{3}}\sigma_y,K_3=\sqrt{\dfrac{p}{3}}\sigma_z$,
\end{center}
where  $p$ is the channel strength with $0 \leq p \leq 1$.

\subsubsection{Phase damping channel}

The action of the phase damping channel on the state $\tilde{\rho}$ can be expressed as \cite{nc}
\begin{equation}
\varepsilon(\tilde{\rho})=\sum_{i} K_i \tilde{\rho} K_{i}^{\dagger}.
\end{equation}
The Kraus operators for Phase damping channel are given by
\[ K_0=\begin{bmatrix}
1&0\\
0&\sqrt{1-p}
\end{bmatrix}
~~and~~ K_1=\begin{bmatrix}
0&0\\
0&\sqrt{p}
\end{bmatrix}\]
Here $p$ is the channel strength with $0 \leq p \leq 1$. 
 \section{The Scenarios}

In order to study the effect of environmental interactions on nonlocal states, we have considered two scenarios as follows:

\begin{figure}[t]
	\centering
	\includegraphics[width=7.5cm,height=4.4cm]{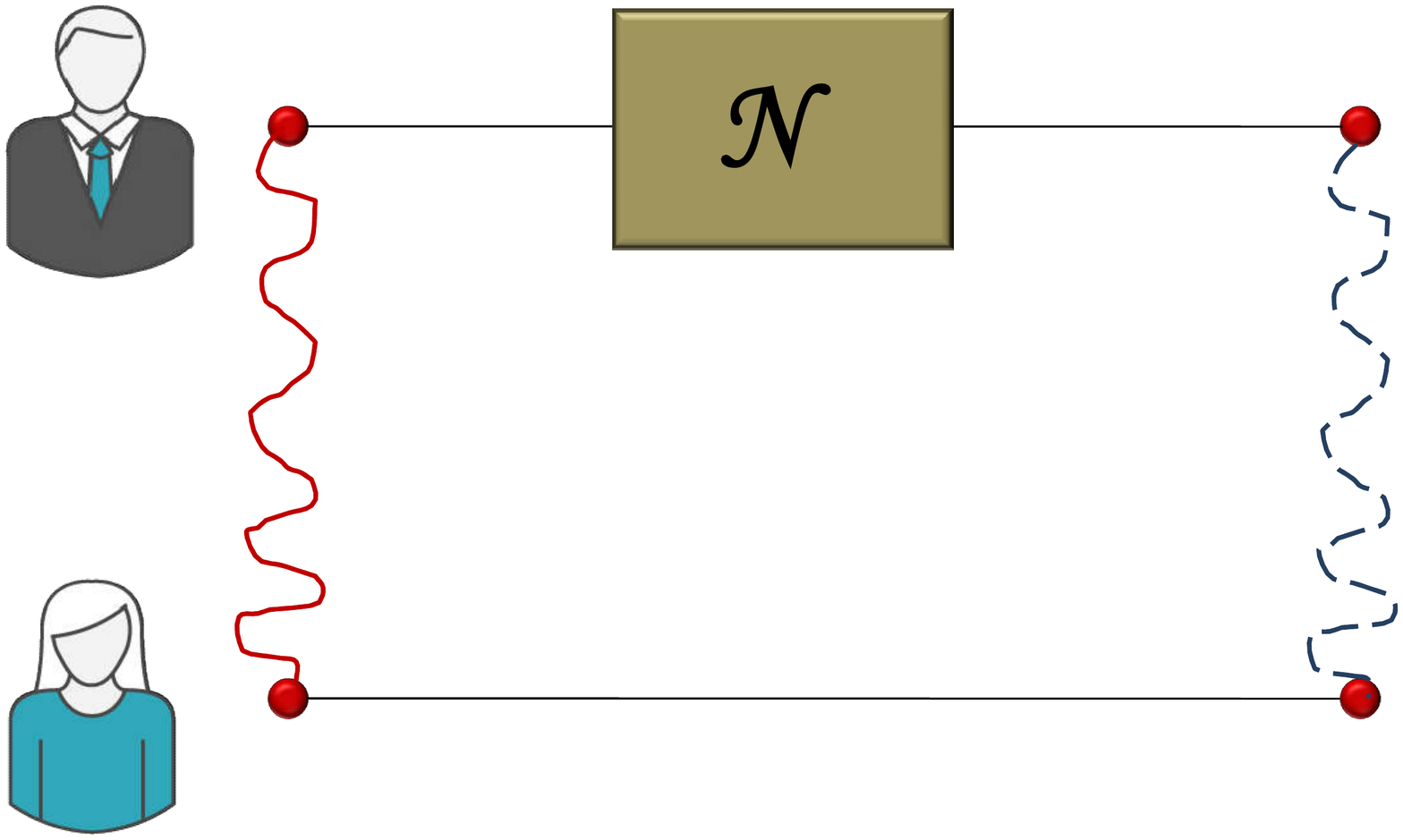}
	\captionof{figure}{Figure depicts single interaction scenario where $\mathcal{N}$ stands for local channel acting on Bob's side.}
	\label{si}
\end{figure}

\subsection{Single interaction}
Consider that Alice and Bob (the set of local measurements performed by Alice is spacelike separated from that of Bob) share a nonlocal two qubit mixed state. Bob's qubit interacts with the environment simulated by a quantum channel of strength $p$. They finally obtain a family of mixed states which are dependent on $p$. We  calculate the range of the channel strengths for which the state becomes Bell-CHSH local, absolutely Bell-CHSH local,absolutely 3-settings unsteerable. Figure \ref{si} depicts the scenario.

\vskip 2cm

\subsection{Double Interaction}

\begin{figure}[t]
	\centering
	\includegraphics[width=7.5cm,height=4.4cm]{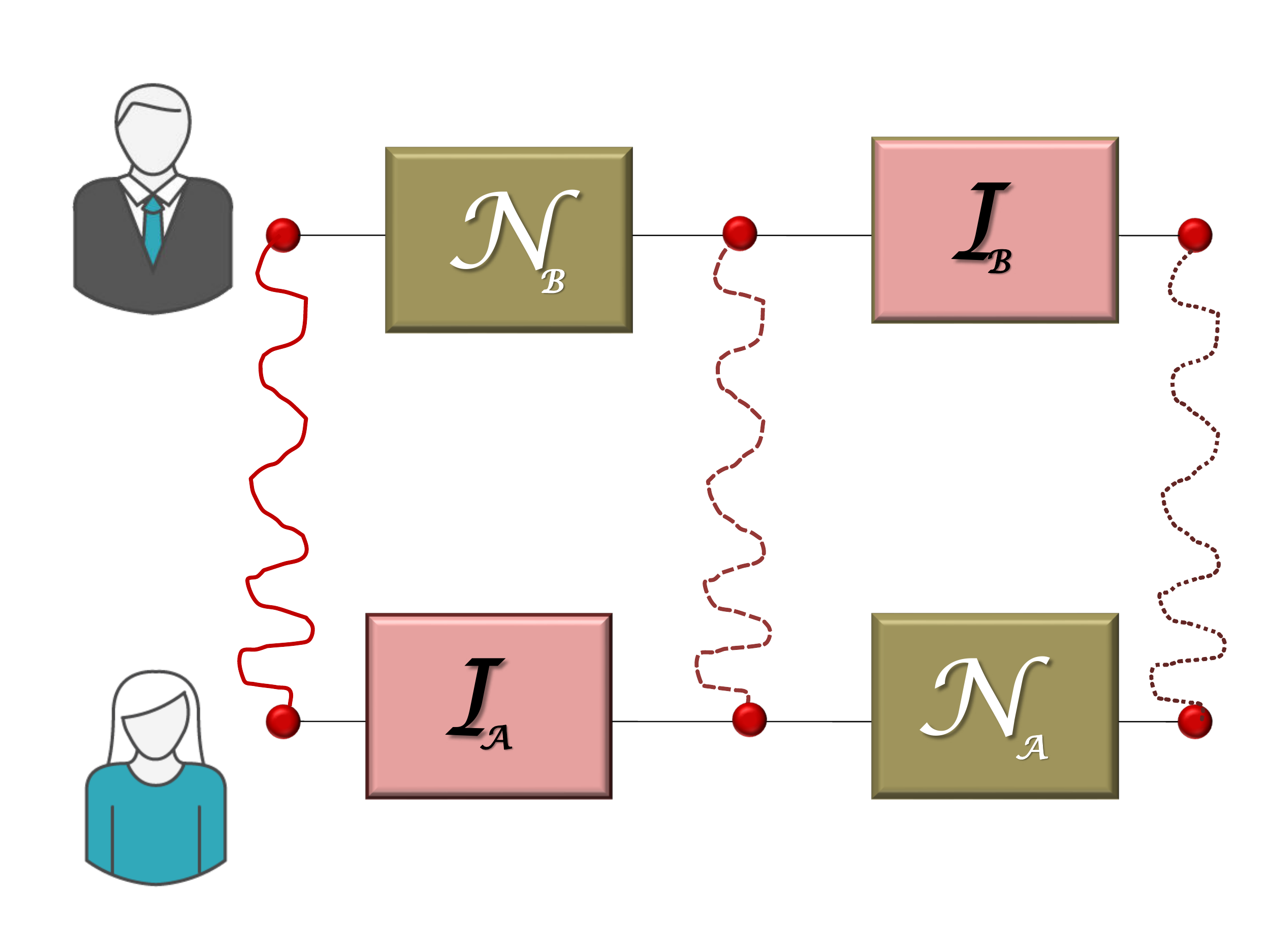}
	\captionof{figure}{ Figure depicts sequential interaction where $\mathcal{N}_B$ and $\mathcal{N}_A$ stand for local channels acting on Bob's and Alice's side respectively. $\mathcal{I}_B$ and $\mathcal{I}_A$ represent identity operations on respective sides.}
	\label{di}
\end{figure}

In this case the situation differs from the above in the sense that now both the qubits interact with environment. The quantum channel is of the same strength $ p $. In this case too we obtain the spread of the channel strength pertaining to the same manifestations as above. The scenario is represented by Figure \ref{di}.\\
In what follows below we present the first significant result of our work.

\section{Nonlocal state $\rightarrow$ Bell-CHSH local state, absolutely Bell-CHSH local state and absolutely 3-settings unsteerable state} 
\begin{figure}[t!]
	\begin{center}
		\begin{framed}
			\resizebox{5cm}{5cm}{\includegraphics{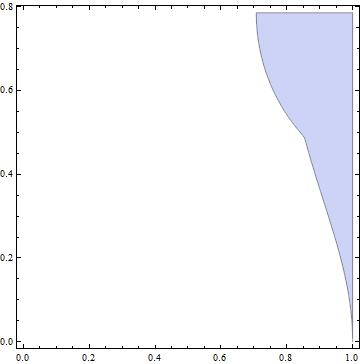}}
			\caption{The horizontal axis represents $\lambda$ and the vertical axis represents $\theta$. The shaded region depicts the nonlocal region of the state (\ref{ins})} \label{gisin}
		\end{framed}
	\end{center}
\end{figure} 
Let us consider that the following two parameter family of mixed states is shared among Alice and Bob where the local measurements performed by Alice are spacelike separated from that of Bob ,
\begin{equation}
\label{ins}
\rho_i(\lambda,\theta)=\begin{bmatrix}
\dfrac{1-\lambda}{2}&0&0&0\\
0&\lambda\sin^{2}\theta&\dfrac{\lambda}{2}\sin 2 \theta&0\\
0&\dfrac{\lambda}{2}\sin 2 \theta&\lambda\cos^{2}\theta&0\\
0&0&0&\dfrac{1-\lambda}{2}
\end{bmatrix}.
\end{equation}
Figure \ref{gisin} depicts the nonlocal region for the family of mixed states given by Eq.(\ref{ins}). Let us choose two initial states from the above two parameter family of states, one with $\lambda=0.95$, $ \theta=0.6$ and another with $\lambda = 0.80$ and $\theta = 0.6$, such that the two initial states are nonlocal. Let us consider that the initial states are subjected to the aforementioned single and double interactions of different quantum channels with channel strength $p$. Let $R_1$, $R_2$ and $R_3$ denote the ranges of $p$ for which the states obtained are Bell-CHSH local, absolutely Bell-CHSH local, and absolutely 3-settings unsteerable, respectively.\\

\begin{widetext}
In what follows below the following two tables represent the various manifestations under different channel parameters for the single and double interactions respectively,

\begin{table}[h]
	\caption{Single interaction} 
	\centering 
	
	\begin{tabular}{c c c c c} 
		\hline 
		\textbf{Channels}  & ~~\textbf{Initial state parameters} & \textbf{$R_1$} & \textbf{$R_2$} & \textbf{$R_3$} \\ [0.2ex] 
		\hline 
		Phase flip & $\lambda=0.95, \theta=0.6$ & [0.1492, 0.8508] & ~[0.2252, 0.7747]~ & -- \\ [0.2ex]
		\hline
		
		Phase flip & $\lambda=0.8, \theta=0.6$ & ~[0.0258, 0.9742]~ & [0.0675,0.9325] & [0.1743, 0.8257] \\
		\hline
		
		Bit flip & $\lambda=0.95, \theta=0.6$ & [0.2445,0.7342]~ & ~[0.3532,0.6443]~ & [0.3807,0.6193] \\
		\hline
		Bit flip & $\lambda=0.8, \theta=0.6$ & [0.0531,0.9468]~ & ~[0.1250, 0.8750]~ & [0.2, 0.8] \\
		\hline
		Depolarizing  & $\lambda=0.95, \theta=0.6$ & [0.0685, 1]~ & ~[0.1928,1]~ & ~[0.2893,1] \\ 
		\hline 
		Depolarizing & $\lambda=0.8, \theta=0.6$ & [0.0692,1]~ & ~[0.1928, 1]~ & ~[0.2893,1] \\
		\hline
		Phase damping & $\lambda=0.95, \theta=0.6$ & [0.3723,1]~ & ~[0.7846, 1]~ & -- \\
		\hline
		Phase damping & $\lambda=0.8, \theta=0.6$ & [0.0516,1]~ & ~[0.1350, 1]~ & ~[0.3485, 1] \\

		\hline 
	\end{tabular}
\end{table}

\begin{table}[h]
	\caption{Double interaction} 
	\centering 
	
	\begin{tabular}{c c c c c} 
		\hline 
		Channels  & ~~Initial state parameters & $R_1$ & $R_2$ & $R_3$ \\ [0.3ex] 
		\hline 
		Phase flip & $\lambda=0.95, \theta=0.6$ & [0.1492, 0.8508] & ~[0.2252, 0.7747]~ & -- \\ 
		\hline
		Phase flip & $\lambda=0.8, \theta=0.6$ & ~[0.0131, 0.9869]~ & [0.3050,0.9650] & [0.0964, 0.9036] \\
		\hline
		Bit flip & $\lambda=0.95, \theta=0.6$ & [0.1378,0.8622]~ & ~[0.1856,0.8144]~ & [0.2256,0.7744] \\
		\hline
		Bit flip & $\lambda=0.8, \theta=0.6$ & [0.0273,0.9727]~ & ~[0.0654, 0.9345]~ & [0.1093, 0.8907] \\
		\hline
		Depolarizing  & $\lambda=0.95, \theta=0.6$ & [0.0354, 1]~ & ~[0.0727,1]~ & ~[0.1560,1] \\ 
		\hline 
		Depolarizing & $\lambda=0.8, \theta=0.6$ & [0.0196,1]~ & ~[0.0481, 1]~ & ~[0.0922,1] \\
		\hline
		Phase damping & $\lambda=0.95, \theta=0.6$ & [0.2077,1]~ & ~[0.5359, 1]~ & -- \\
		\hline
		Phase damping & $\lambda=0.8, \theta=0.6$ & [0.0261,1]~ & ~[0.0699, 1]~ & ~[0.1928, 1] \\

		\hline 
	\end{tabular}
\end{table}

In tables I and II we have shown the effect of single and sequential interaction of different channels with respect to two different initial states. 

\end{widetext}
It is significant to note that for {\it bit-flip} and {\it phase-flip} channels, along with the lower bound there is also an upper bound on the noise parameter ($p$), between which they act as resource destroying maps. This counter-intuitive feature has a deep connection with the structure of the state (\ref{ins}). Single interaction of the {\it phase flip} noise makes $(\lambda,\theta)\to(\lambda,-\theta)$, whereas the {\it bit flip} noise makes a block relabelling of the initial quantum state. As a result, for the value of noise parameter beyond an upper-bound, the resource destroying power of the map is decreased in case of the above mentioned channels.
It is also worthwhile to note that barring the first row and one range in the second row the ranges in Table I are subsets of their counterpart in Table II. This agrees with intuition as Table II depicts the range pertaining to double interaction.

\section{Generating entangled states admitting LHS model from initially nonlocal state} 
There has been considerable work in recent times on the construction of states admitting LHV and LHS models using techniques from semidefinite programming \cite{caval,brunner}. In what follows below, we lay down two scenarios which generate entangled  states admitting LHS model from initially nonlocal state.\\

\textit{\textbf{The scenario using phase damping channel:}}\\

Let us consider the following two parameter family of bipartite two-qubit mixed quantum states initially shared between Alice and Bob (where the local measurements performed by Alice are spacelike separated from that of Bob),
\begin{widetext}
\begin{equation}
\label{is}
\rho_i(q,s)=q \Big( s \vert \phi^{+}\rangle \langle\phi^{+}\vert + (1-s)\dfrac{\mathbb{I}}{4} \Big) + (1-q) \Big( \dfrac{1}{2}\vert 00 \rangle \langle 00\vert+\dfrac{1}{2}\vert11 \rangle \langle 11\vert \Big),
\end{equation}
where, $|0\rangle$ and $|1\rangle$ are the eigenstates of the operator $\sigma_z$ with eigenvalues $+1$ and $-1$ respectively, $| \phi^{+} \rangle = \dfrac{1}{\sqrt{2}} \Big( |00 \rangle + | 11 \rangle \Big)$, $\mathbb{I}$ is the identity operator, $0\leq q \leq 1$ and $0 \leq s \leq 1$.

\end{widetext}

Now, we use the phase damping channel . Let us choose $q=0.96$, $s=0.74$ such that the initial state given by Eq.(\ref{is}) is nonlocal. In case of the phase damping channel with channel strength $p$, if this state undergoes single interaction as described earlier, then for $p=p_1=0.65$, the state becomes
\begin{equation}
\label{bs}
\rho_f=\dfrac{1}{2}\sigma + \dfrac{1}{2}\Big( \dfrac{1}{2}\vert 00 \rangle\langle00\vert+\dfrac{1}{2}\vert11\rangle\langle11\vert \Big),
\end{equation}
where $\sigma$ is the two qubit isotropic state given by $\sigma = \dfrac{1}{2} \Big( | \phi^{+} \rangle \langle \phi^{+}| + \dfrac{\mathbb{I}}{4} \Big)$, $| \phi^{+} \rangle = \dfrac{1}{\sqrt{2}} \Big( |00 \rangle + | 11 \rangle \Big)$.
It has been shown that the correlation produced by the state $\rho_f$ given by Eq.(\ref{bs}) has a local hidden variable-local hidden state (LHV-LHS) model \cite{bs}. Hence, the correlation produced by the state $\rho_f$ has a LHV model, since the states having a LHV-LHS model form a subset of the states having a LHV model.
It can easily be checked that the state $\rho_f$ is absolutely Bell-CHSH local according to the condition given by (\ref{ablc}). The state $\rho_f$ is absolutely 3-settings unsteerable also according to the condition given by (\ref{ausc}).

Again, in case of the phase damping channel with channel strength $p$, if the state $\rho_i(q,s)$ given by Eq.(\ref{is}) (with $q=0.96$, $s=0.74$) undergoes sequantial interaction as described earlier, then for $p=p_2=0.41$ the state becomes $\rho_f$ given by Eq.(\ref{bs}).

Hence, here we have presented the transformation of a nonlocal state into an absolutely Bell-CHSH local state (as well as absolutely 3-settings unsteerable state) with an LHV model as well as LHS model under single and double interaction of the phase damping channel.
It is clear that $p_2 < p_1$, which implies that the state $\rho_i(q,s)$ (with $q=0.96$, $s=0.74$) can be transformed into an absolutely Bell-CHSH local state (as well as an absolutely 3-settings unsteerable state) having a LHV model  as well as LHS model under double interaction of the phase damping channel with smaller channel strength compared to that under single interaction.\\

\textit{ \textbf{The scenario using depolarizing channel:}}\\

 In this case let us choose $q=0.34$, $s=0.97$ such that the initial state given by Eq.(\ref{is}) is nonlocal. In case of the depolarizing channel with channel strength $p$, if this state undergoes single interaction as described earlier, then for $p=p_1=0.18$ the state becomes $\rho_f$ given by Eq.(\ref{bs}).
If the state $\rho_i(q,s)$ given by Eq.(\ref{is}) (with $q=0.34$, $s=0.97$)  undergoes sequential interaction of the depolarizing channel as described earlier, then for  $p=p_2=0.10$ the state becomes $\rho_f$ given by Eq.(\ref{bs}).

Hence, here we have presented the transformation of a nonlocal state into an absolutely Bell-CHSH local state (as well as absolutely 3-settings unsteerable state) having a LHV model as well as LHS model under single and double interaction of the depolarizing channel. Note that since, $p_2 < p_1$, the state $\rho_i(q,s)$ (with $q=0.34$, $s=0.97$) can be transformed into an absolutely Bell-CHSH local state (as well as an absolutely 3-settings unsteerable state) having an LHV model as well as LHS model under double interaction of the depolarizing channel with smaller channel strength compared to that under single interaction.

\section{Steerability breaking channels and Depolarizing map}

In \cite{bs}, authors gave a sufficient criterion for unsteerability pertaining to a state in two qubits. They considered a state in two qubits as given below: 
\begin{equation}
  \chi=\frac{1}{4}[I \otimes I + \vec{a}.\vec{\sigma} \otimes I + \underset{i}{\Sigma} t_{ii} \sigma_i \otimes \sigma_i] , i=1,2,3 
  \label{unsteer}
\end{equation}
 Here $\mathbb{I}$ is the identity operator acting on $\mathbb{C}^2$; $\sigma_i$s are the three Pauli matrices; $\vec{r}, \vec{a}$ is vector in $\mathbb{R}^3$ with norm less than or equal to unity; $\vec{a}.\vec{\sigma} = \sum_{i=1}^{3} a_i \sigma_i$; $t_{ii} = Tr[ \chi (\sigma_i \otimes \sigma_i )]$. The state in Eq.(\ref{unsteer}) is unsteerable if $\underset{\hat{x}}{Max}[(\vec{a}.\hat{x})^2 + 2 \vert \vert T \hat{x} \vert \vert ] \leq 1 $, where $ \vert \vert \cdot \vert \vert $ is the Euclidean vector norm, $T=[t_{ij}]$ and $\hat{x}$ is a unit vector.

Since  $\underset{\hat{x}}{Max}{(\vec{a}.\hat{x})^2}=\vert \vec{a} \vert^{2}$ and $\underset{\hat{x}}{Max}{\vert \vert T \hat{x} \vert \vert} = \sqrt{\lambda_{max}}$, where $\lambda_{max}$ is the largest eigenvalue of $T^{\dagger}T$, we obtain the condition as 
\begin{equation}
 \vert \vec{a} \vert^{2} +   2 \sqrt{\lambda_{max}} \leq 1 
\end{equation}

The depolarizing map is a significant completely positive map in quantum information processing. It has been studied in the context of incompatibility breaking channels \cite{incomp}. Its action on a single qubit system $A$ is given by 
\begin{equation}
    \Gamma_{\epsilon}(A) = \epsilon A + (1- \epsilon).\frac{1}{2}Tr[A]I
\end{equation}
where $\epsilon$ is the channel parameter. Now, consider the state given in Eq.(\ref{unsteer}). Then due to the action of the depolarizing map on the state we obtain,
\begin{equation}
    ( \Gamma_{\epsilon} \otimes I) \chi = \frac{1}{4}[I \otimes I + \epsilon(\vec{a}.\vec{\sigma})+ \underset{i}{\Sigma} \epsilon  t_{ii} \sigma_i \otimes \sigma_i]
\end{equation}
On using the unsteerability criterion, the resultant state is unsteerable if,
\begin{equation}
    \vert \epsilon \vec{a} \vert^2 + 2 \epsilon \sqrt{Max(t_{11}^2,t_{22}^2,t_{33}^2)} \leq 1
\end{equation}
This gives us a bound on the parameter $\epsilon$ for which the map becomes steerability breaking. One may also consider the map to be incompatibility breaking \cite{incomp}, as steerability and incompatibility of measurements have a one to one correspondence \cite{guhnesteering}. However, one must note that since the unsteerability criterion is only sufficient, the bounds are not tight.

\section{Discussion and Conclusions} 
In the present work we have discussed the possible detrimental effects of the environment on nonlocal resources. In the context of any quantum information protocol one cannot ignore environmental influences. In the present work, we have documented various scenarios where nonlocality is lost, sometimes to the extent that it cannot be retrieved even with the strong resource of global unitary operations.  We have illustrated various channel actions which destroy certain correlations. From the perspective of experiments, one should avoid those maps to preserve various nonlocal correlations. It is in this line of thought, that our work assumes significance. Precisely we show that there are instances where a quantum state loses its ability to violate the Bell-CHSH inequality or the 3-settings steering inequality. Our submission contributes to the notion of nonlocality breaking channels and also focusses on steerability breaking maps.  

Our work also is an attempt towards understanding the notion of absolutely local maps which render the resultant state useless regarding nonlocality. This is akin to the work already done in entanglement theory where absolute separability maps were considered \cite{assd}. In the tables presented,as expected intuitively, we observe that in most of the cases the parameter ranges obtained for a single interaction are subsets of that obtained for double interaction. We would like to reiterate that our work concerns mixed entangled states as pure states always violate some suitably chosen Bell's inequality. 

In the present work we have laid down a theoretical proposal to generate entangled states admitting LHS model. As such entangled states occupy an intriguing existence in the theory of nonlocality and foundations of quantum theory, one might consider the experimental demonstration of the result.
We  have obtained bounds on the parameter of the depolarizing map for which it becomes steerability breaking or one may also term them as incompatibility breaking. This does not depend on any particular inequality.
This work also leads to certain open questions and possibilities for subsequent studies. For example, one may probe other channel parameters for which they become steerability breaking. An extension of the work in higher dimensions and multipartite systems also deserve attention.

 \section*{Acknowlegement}

We would like to gratefully acknowledge fruitful discussions with Prof. Guruprasad Kar. BB acknowledges support from the DST-INSPIRE, Government of India. DD acknowledges the financial support from University Grants Commission (UGC), Government of India. ASM acknowledges Project no. DST/ICPS/Qust/2018/98 of Department of Science and Technology, Government of India. NG would like to acknowledge support from the Research Initiation Grant of BITS-Pilani, Hyderabad vide letter no.
BITS/GAU/RIG/2019/H0680 dated 22nd April, 2019.

\end{document}